\documentclass[prb,twocolumn,showpacs,preprintnumbers,amsmath,amssymb,superscriptaddress]{revtex4}

\usepackage{graphicx}% Include figure files
\usepackage{dcolumn}% Align table columns on decimal point
\usepackage{bm}% bold math
\begin{document}
\title{
Inhomogeneous Gutzwiller approximation with random phase fluctuations
for the Hubbard model}
\author{G. Seibold}
\affiliation{Institut f\"ur Physik, BTU Cottbus, PBox 101344,
         03013 Cottbus, Germany}
\author{F. Becca}
\affiliation{Institut de Physique Th\'eorique, Universit\'e de
Lausanne, CH-1015 Lausanne, Switzerland}
\author{J. Lorenzana}
\affiliation{Consejo Nacional de Investigaciones Cient\'ificas y
T\'ecnicas and Instituto Balseiro, Centro At\'omico Bariloche,
8400 S. C. de Bariloche, Argentina}
\date{\today}

\begin{abstract}
We present a detailed study of the time-dependent Gutzwiller
approximation for the Hubbard model.
The formalism, labelled
GA+RPA, allows us to compute random-phase approximation-like (RPA)
fluctuations on top of the Gutzwiller approximation (GA).  No
restrictions are imposed on the charge and spin configurations
which makes the method suitable for the calculation of linear
excitations around symmetry-broken solutions.
Well-behaved sum rules are obeyed as in the Hartree-Fock (HF)
plus RPA approach. Analytical results for a two-site model and
numerical results for charge-charge and current-current dynamical
correlation functions in one and two dimensions are compared with
exact and HF+RPA results, supporting the much better
performance of GA+RPA with respect to conventional HF+RPA theory.
\end{abstract}


\maketitle

\section{Introduction}
The Gutzwiller variational wave function together with the
Gutzwiller approximation (GA)~\cite{GUTZ} is a widely used
approach in order to deal with Hubbard-type models. Originally
introduced in order to explore the possibility of ferromagnetism
within the Hubbard model (see e.g., Ref.~\onlinecite{BUENEMANN}
and references therein) its popularity resides in the fact that it
captures correlation effects like the band narrowing already on
the variational level. More recently the GA has been also used for
realistic band structure
computations.~\cite{BUENEMANN,bue98,zei95} Since in the Hubbard
model one has a competition between delocalization, from the
hopping of the charge carriers, and localization, from the onsite
interaction $U$, the idea is to apply a projector to a given
Slater determinant which reduces the number of doubly occupied
sites. Within the GA one has to minimize an energy functional
which is composed of a renormalized kinetic term and the
interaction energy $UD$, where $D$ denotes the concentration of
doubly occupied sites.

On the other hand, mean-field theories, like Hartree-Fock (HF),
are usually only the first step in a many-body computation and it
is often desirable to include the effect of fluctuations within
the random-phase approximation (RPA). In case of HF this has been
achieved by numerous techniques (for an overview see e.g.,
Ref.~\onlinecite{PINES}), however, the development of a similar
scheme in the GA has been a long-standing problem  of the
condensed-matter many-body community. The major step in this
direction was the reformulation of the GA by Kotliar and
Ruckenstein (KR) within the so-called four slave-boson
approach.~\cite{KR} This method maps the physical hole (or
particle) into products of fermion and boson operators where the
latter additionally label the occupancy of the site. At the
saddle-point level the bosons are replaced by their mean-field
values and one recovers the GA energy functional showing its
underlying mean-field character.

The KR slave-boson formulation offers the possibility of going
beyond the Gutzwiller result by the inclusion of transversal spin
degrees of freedom.~\cite{WH} Moreover, in principle, it provides
a controlled scheme of including fluctuations beyond the
mean-field solution. However, the expansion of the KR hopping
factor $z^{SB}$ is a highly nontrivial task, both with respect to
the proper normal ordering of the bosons and with respect to the
correct continuum limit of the functional
integral.~\cite{ARRIGONI} Expansions around the slave-boson saddle
point have been performed for homogeneous systems in
Refs.~\onlinecite{LI,LAVAGNA} in order to calculate correlation
functions in the charge and longitudinal spin channels.
Furthermore, the optical conductivity in the paramagnetic regime
of the Hubbard model has been calculated in
Ref.~\onlinecite{RAIMONDI1}. A severe difficulty in this approach
is the fact that the KR choice for the hopping factor did not lead
to controlled sum rules.~\cite{RAIMONDI2} Moreover, to our
knowledge, this approach has not been extended to symmetry broken
states due to the complexity of the computation.

Recently, two of us have presented a computation of RPA
fluctuations on top of GA states (GA+RPA).~\cite{GOETZ3} Our
approach borrows ideas from well developed techniques in nuclear
physics,~\cite{THOULESS} and RPA fluctuations are obtained in the
small oscillation limit of a time-dependent Gutzwiller
approximation. Since response functions are derived for systems
with completely unrestricted charge and spin distributions GA+RPA
is suitable also for the calculation of charge excitations on
solutions with inhomogeneous textures. A key point of the GA+RPA
approach is the proper determination of the time dependence of the
variational double occupancy parameter. We have adopted an
antiadiabatic approximation in the sense that the double occupancy
adjusts instantaneously to the time evolution of the single
particle densities. In this context our approach can be viewed as
a generalization of the Fermi liquid analysis of
Vollhardt.~\cite{VOLLHARDT}

In this paper we use the  GA+RPA to compute various correlation
functions in the one-band Hubbard model and compare with exact
diagonalization and HF+RPA results.

This paper is organized as follows.
In Sec.~\ref{model} we present the formalism. We concentrate on the
case of the one-band Hubbard model although generalization
to more complicated models is straightforward.
From the expansion of the GA
energy functional up to second order in the densities we
demonstrate how the RPA response functions can be calculated and
demonstrate that standard sum rules are obeyed. The method is
illustrated in Sec.~\ref{two} for the two-site Hubbard model which
can be treated analytically. Finally, in Sec.~\ref{dynamics} we
compare the GA+RPA excitation spectra with exact diagonalization
and HF+RPA results respectively.

\section{Model and Formalism}
\label{model}
We consider the one-band Hubbard model
\begin{equation}\label{HM}
H=\sum_{ij,\sigma}t_{ij} c_{i,\sigma}^{\dagger}c_{j,\sigma} + U\sum_{i}
n_{i,\uparrow}n_{i,\downarrow},
\end{equation}
where $c_{i,\sigma}^{(\dagger)}$ destroys (creates) an electron
with spin $\sigma$ at site
i, and $n_{i,\sigma}=c_{i,\sigma}^{\dagger}c_{i,\sigma}$. $U$ is the
onsite Hubbard repulsion and $t_{ij}$ denotes the hopping parameter between
sites $i $ and $j$.

\subsection{Gutzwiller approximation}
In its original formulation, the GA yields an approximation for
the energy of a uniform paramagnetic system.~\cite{GUTZ} Only in
the late 80's this approach has been consistently generalized to
an unrestricted Slater determinant within the Kotliar and
Ruckenstein slave-boson approach.~\cite{KR} The same unrestricted
Gutzwiller energy functional has been obtain by
Gebhard~\cite{geb90} exploiting the fact that the GA becomes the
exact solution of the Gutzwiller variational problem in the limit
of infinite spatial dimensions.~\cite{METZNER}

In Gebhard's formulation the variational wave function is written
as:\cite{geb90,zei95}
\begin{eqnarray}
  \label{eq:gwf}
|\Psi\rangle&=&\prod_i \frac {\hat U_i}{K_i^{1/2}}|SD\rangle,\\
 \hat U_i&=&\exp
\left( - \gamma_i n_{i,\uparrow}n_{i,\downarrow}- \sum_{\sigma}
 \mu_{i,\sigma} n_{i,\sigma}\right)
\end{eqnarray}

where $K_i=\langle\Psi |{\hat U_i}{\hat U_i}|\Psi\rangle$. In Eq.
(\ref{eq:gwf})  $|SD\rangle$ denotes a Slater determinant which
already incorporates the Hartree contribution of the local
interactions and which has to be determined variationally. The
solution of the variational problem in the limit of infinite
dimensions turns out to be the GA generalized for an arbitrary
charge and spin distribution of the SD. The $\mu_{i,\sigma}$ act
as local chemical potentials and are determined within the GA by
the infinite dimension prescription that the diagonal charges are
not renormalized:

$$\langle\Psi |n_{i,\sigma}|\Psi\rangle=\langle SD
|n_{i,\sigma}|SD\rangle.$$

As a result one obtains for the GA energy
functional:~\cite{KR,geb90}
\begin{eqnarray}\label{EGW}
&&E^{GA}[\rho,D]=\sum_{ij,\sigma}t_{ij}
z^{GA}_{i,\sigma}z^{GA}_{j,\sigma} \rho_{ji,\sigma}
+U\sum_{i}D_{i} \label{E1},\\
&&z^{GA}_{i,\sigma}=\frac{\sqrt{(1-\rho_{ii}+D_i)(\rho_{ii,\sigma}-D_i)}
+\sqrt{D_i(\rho_{ii,-\sigma}-D_i)}}{\sqrt{
\rho_{ii,\sigma}(1-\rho_{ii,\sigma})}}, \nonumber
\end{eqnarray}
which is a functional of the  density matrix
$\rho_{ij,\sigma}\equiv \langle SD| c_{j,\sigma}^{\dagger}
c_{i,\sigma} |SD\rangle$ and the double occupancy parameters
$D_i$. We  denote the set of all matrix elements
$\{\rho_{ij,\sigma} \}$ and $\{ D_i \}$ by  $\rho$ and $D$. Notice
that in this paper we do not consider spin-canted solutions which
would have density matrix elements $\langle SD|
c_{j,\sigma}^{\dagger} c_{i,\sigma'} |SD\rangle \neq 0$ for
$\sigma \neq \sigma'$. However, transverse spin degrees of freedom
can be straightforwardly incorporated within the spin-rotationally
invariant slave boson formulation.~\cite{WH}

We will denote by $|\Psi_0\rangle$ the particular wave function of
form Eq.~(\ref{eq:gwf}) that minimizes the energy. In order to
obtain the associate stationary solution ${\rho^{(0)},D^{(0)}}$
one has to minimize $E^{GA}$ with respect to the double occupancy
parameters $D$ and the density matrix $\rho$, where the latter
variation has to be constrained to the subspace of Slater
determinants by imposing the projector condition
$\rho^2=\rho$:~\cite{RING,BLAIZOT}
\begin{equation}\label{EGW1}
\delta\{E^{GA}[\rho,D]-\mbox{tr}[\Lambda(\rho^2-\rho)]\}=0,
\end{equation}
where $\Lambda$ denotes the Lagrange parameter matrix. It is
convenient to define a Gutzwiller Hamiltonian:~\cite{RING,BLAIZOT}
\begin{equation}
  \label{eq:hgw}
  h_{ij\sigma}[\rho,D]=\frac{\partial E^{GA} } {\partial \rho_{ji\sigma}},
\end{equation}
which is also a functional of $\rho$ and ${D}$.
Variation of Eq.~(\ref{EGW1}) with respect to the density matrix leads to:
\begin{equation}
  h-\rho\Lambda-\Lambda\rho+\Lambda=0.
\end{equation}
The Lagrange parameters can be eliminated~\cite{BLAIZOT} and
together with the variation with respect to $D$ we obtain the self
consistent GA equations:
\begin{eqnarray}\label{eq:ga}
  [h,\rho]&=&0, \\
  \frac{\partial E^{GA}}{\partial D_i}&=&0. \label{eq:gamin}
\end{eqnarray}
The first equation can be solved by diagonalizing both
the Gutzwiller Hamiltonian and the density matrix
by a linear transformation of
the single-particle orbital basis:
\begin{equation}\label{TRAFO1}
c_{i,\sigma}=\sum_{\nu}\psi_{i,\sigma}(\nu)a_{\nu},
\end{equation}
leading to $h^{0}_{\mu\nu}=\delta_{\mu\nu} \epsilon_\nu$. Moreover,
the diagonalized density matrix $\rho \equiv \rho^{(0)}$
has eigenvalue 1 below the Fermi level and eigenvalue 0 above it.
We use Greek letters to denotate any state of this
particular basis and
the nought indicates evaluation in the saddle-point. Additionally  we
denotate states below the Fermi level as hole ($h$) states and the
states above the Fermi level as particle states ($p$).

Notice that
in this base $(\rho^{(0)})^2=\rho^{(0)}$  is trivially
satisfied. $\rho^{(0)}$ acts as a projector onto the hole
states of the saddle-point Slater determinant in the space of the
density matrices whereas
$\sigma^{(0)} \equiv 1-\rho^{(0)}$ corresponds to the projector onto particle
states.

The diagonalization of Eq.~(\ref{eq:ga}) has to be supplemented by
the minimization of the Gutzwiller energy with respect to the
double occupancy parameters of Eq.~(\ref{eq:gamin}). A convenient
method for this purpose in order to obtain inhomogeneous GA
solutions has been discussed in Ref.~\onlinecite{GOETZ1}. Notice
that the unrestricted variational procedure with respect to charge
and (or) spin degrees of freedom prevents the occurrence of the
Brinkmann-Rice transition towards localization~\cite{BR} which has
already been shown in Ref.~\onlinecite{SHIBA} for Ne\'{e}l-type
antiferromagnetism.

\subsection{Derivation of the RPA equation}
Before starting our analysis it is convenient for later use to
define the GA effective operator:
\begin{equation}\label{eq:gaop}
O^{GA}=\sum_{ij\sigma} (o_{ij,\sigma}^{GA} c_{i\sigma}^{\dagger}
c_{j\sigma} +h.c.),
\end{equation}
where $o_{ij\sigma}^{GA}=q_{ij\sigma}o_{ij,\sigma}$ and
$q_{ij\sigma}=1$ if $i=j$ and
$q_{ij\sigma}=z^{GA}_{i,\sigma}z^{GA}_{j,\sigma}$ otherwise.

In order to derive the RPA equation we introduce a small time-dependent
external field added to Eq.~(\ref{HM}):
\begin{equation}\label{eq:fdt}
F(t)=
\sum_{ij\sigma} (f_{ij,\sigma}(t)
c_{i\sigma}^{\dagger} c_{j\sigma} +h.c.),
\end{equation}
with $f_{ij,\sigma}(t)=f_{ij,\sigma}(0) e^{-i\omega t}$.
As a consequence $|\Psi\rangle$,
$|SD\rangle$, and the variational  parameters acquire a time dependence
and an additional term appears in the  energy functional Eq.~(\ref{EGW}):
\begin{eqnarray}\label{eq:fred}
E_f^{GA}[\rho,D](t)&=&\langle\Psi(t)|H_f(t)|\Psi(t)\rangle\nonumber\\
&=&\sum_{ij\sigma}
 (f_{ij\sigma}^{GA}  \rho_{ji,\sigma}  e^{-i\omega t} + h.c.),
\end{eqnarray}
where $f_{ij\sigma}^{GA}=q_{ij\sigma} f_{ij,\sigma}$.

The time dependent field induces small amplitude oscillations
of $D$ and $\rho$ around the GA saddle-point:
\begin{eqnarray}
D &=& D^{(0)} + \delta D(t), \\
\rho &=& \rho^{(0)} + \delta \rho(t).
\end{eqnarray}
The density and double occupancy fluctuations
are constrained by the following requirements:

{\it i}) At all times $\rho$ is constrained to be
the one-body density matrix associated with a Slater
determinant. This can be achieved by imposing:
\begin{equation}\label{eq:sd}
  \rho=\rho^2.
\end{equation}

{\it ii}) The double occupancy is assumed to have a much faster
dynamics than the density matrix so that it can be treated
antiadiabatically. As a consequence, $\delta D$
adjusts instantaneously to the evolution of the density matrix
via the condition
\begin{equation}\label{AAC}
  \frac{\partial E^{GA}[\rho,D]}{\partial \delta D_i}=0.
\end{equation}

In fact Eq.~(\ref{AAC}) constitutes the basic hypothesis of the
present formalism which is necessary in order to derive
an effective {\it Gutzwiller interaction} between particles (see below).
We expect this approximation to be accurate for sufficiently
low-energy excitations. At high energies one can check the accuracy
and the limits of validity of this approximation by
comparing with exact diagonalization, as done in the following
sections. Surprisingly, it turns out to be accurate at least up to
energies of the order of the Mott-Hubbard gap.

As in any small amplitude approximation, we start by expanding
the GA energy [Eqs.~(\ref{EGW}) and~(\ref{eq:fred})] around the saddle-point.
The first part, Eq.~(\ref{EGW}), is needed up to
second order in the density and double occupancy deviations:
\begin{eqnarray}
E^{GA}[\rho,D]&=&E_0+ \mbox{tr}(h^0\delta{\rho}) \nonumber \\
&+&\sum_{ij,\sigma}t_{ij}
\lbrack z^{GA}_{i,\sigma} \delta_1 z^{GA}_{j,\sigma}
+ z^{GA}_{j,\sigma} \delta_1 z^{GA}_{i,\sigma}\rbrack
\delta \rho_{ji,\sigma} \nonumber \\
&+&\sum_{ij,\sigma}t_{ij} \rho_{ji,\sigma}
\delta_1 z^{GA}_{i,\sigma}\delta_1 z^{GA}_{j,\sigma} \nonumber \\
&+&\sum_{ij,\sigma}t_{ij} \rho_{ji,\sigma} \lbrack
z^{GA}_{i,\sigma} \delta_2 z^{GA}_{j,\sigma}
+ z^{GA}_{j,\sigma} \delta_2 z^{GA}_{i,\sigma}\rbrack. \label{EGW2}
\end{eqnarray}
Here $E_0$ denotes the saddle-point (mean-field) energy and the
trace includes sum over spins.
We have used the following abbreviations for the $z$-factor expansion
\begin{eqnarray}
\delta_1 z^{GA}_{i,\sigma} &\equiv&
\frac{\partial z^{GA}_{i,\sigma}}{\partial D_i}
\delta D_i + \sum_{\sigma'}
\frac{\partial z^{GA}_{i,\sigma}}{\partial \rho_{ii,\sigma'}}
\delta \rho_{ii,\sigma'},  \label{ZZ1} \\
\delta_2 z^{GA}_{i,\sigma} &\equiv&
\frac{1}{2} \frac{\partial^2 z^{GA}_{i,\sigma}}{\partial D_i^2}
(\delta D_i)^2 + \sum_{\sigma'}
\frac{\partial^2 z^{GA}_{i,\sigma}}{\partial D_i \partial \rho_{ii,\sigma'}}
\delta D_i \delta \rho_{ii,\sigma'} \nonumber \\
&+& \frac{1}{2} \sum_{\sigma'\sigma''}
\frac{\partial^2 z^{GA}_{i,\sigma}}
{\partial \rho_{ii,\sigma'}\partial \rho_{ii,\sigma''}}
\delta \rho_{ii,\sigma'}\delta \rho_{ii,\sigma''}. \label{ZZ2}
\end{eqnarray}
To proceed further it is convenient to cast the second order
expression in matrix form
\begin{eqnarray}\label{eq:e2}
E^{GA}[\rho,D]&=&E_0+ \mbox{tr}(h^0\delta{\rho}) + \frac{1}{2}
\delta\rho_{ji,\sigma} L_{ijkl}^{\sigma\sigma'}
\delta\rho_{lk,\sigma'} \nonumber \\
&+&\frac{1}{2} \delta D_i K_{ij} \delta D_j + \delta D_i S_{i,kl\sigma}
\delta \rho_{lk\sigma},
\end{eqnarray}
where the matrix multiplications imply the Einstein sum convention
and the definitions for the matrices $L$, $K$ and $S$ follow immediately
from Eqs.~(\ref{EGW2}),~(\ref{ZZ1}) and~(\ref{ZZ2}). The nonzero matrix
elements are given by:
\begin{eqnarray} \label{eq:lks}
  L_{ii,ii}^{\sigma\sigma'}&=&\sum_{j\sigma''}t_{ij}
\frac{\partial^2 z_{i,\sigma''}}
{\partial\rho_{ii,\sigma}\partial\rho_{ii,\sigma'}} z_{j,\sigma''}
(\rho_{ij,\sigma''}+\rho_{ji,\sigma''}), \nonumber\\
 L_{ii,jj}^{\sigma\sigma'}&=&\sum_{\sigma''}t_{ij}
\frac{\partial z_{i,\sigma''}}{\partial\rho_{ii,\sigma}}
\frac{\partial z_{j,\sigma''}}{\partial\rho_{jj,\sigma}}
(\rho_{ij,\sigma''}+\rho_{ji,\sigma''})\ \ \ \ i\ne j, \nonumber\\
L_{ii,ij}^{\sigma\sigma'}&=&L_{ij,ii}^{\sigma'\sigma}=t_{ij}\frac{\partial z_{i,\sigma`}}{\partial\rho_{ii,\sigma}} z_{j,\sigma'}\ \ \ \ i\ne j, \nonumber\\
K_{ii}&=&\sum_{j\sigma}t_{ij}
\frac{\partial^2 z_{i,\sigma}}
{D_i^2} z_{j,\sigma} (\rho_{ij,\sigma}+\rho_{ji,\sigma}), \\
K_{ij}&=&\sum_{\sigma}t_{ij}
\frac{\partial z_{i,\sigma}}{\partial D_i}
\frac{\partial z_{j,\sigma}}{\partial D_j}(\rho_{ij,\sigma}+\rho_{ji,\sigma})\ \ \ \ i\ne j, \nonumber\\
S_{i,ii\sigma}&=&\sum_{j\sigma'}t_{ij} \frac{\partial^2 z_{i,\sigma'}}
{\partial D_i\partial\rho_{ii,\sigma}} z_{j,\sigma'}
(\rho_{ij,\sigma'}+\rho_{ji,\sigma'}), \nonumber\\
S_{i,jj\sigma}&=&t_{ij}\sum_{\sigma'}\frac{\partial z_{i,\sigma'}}
{\partial D_i} \frac{\partial z_{j,\sigma'}}
{\partial \rho_{jj,\sigma}} (\rho_{ij,\sigma'}+\rho_{ji,\sigma'})\ \ \ \ i\ne j, \nonumber\\
S_{i,ij\sigma}&=& t_{ij}
 \frac{\partial z_{i,\sigma}}{\partial D_i} z_{j,\sigma}\ \ \ \ i\ne j. \nonumber
\end{eqnarray}
Notice the formal similarity between Eq.~(\ref{eq:e2}) and
an electron-boson problem where particle-hole excitations interact with a
bosonic degree of freedom in the place of $D$. The matrix $K$ plays the role
of a double occupancy stiffness and $S$ a double occupancy-electron
interaction.

We can integrate out the $D$
fluctuations using the antiadiabaticity condition Eq.~(\ref{AAC}).
First, we express $\delta D_i$ in terms of the density fluctuations via
\begin{equation}\label{eq:ddr}
\delta D_i = - (K^{-1})_{ij}S_{j,kl\sigma}\delta\rho_{lk,\sigma},
\end{equation}
which finally yields an expansion of the
energy as a functional of $\delta \rho$ alone
$\widetilde E[\rho]\equiv E^{GA}[\rho,D(\rho)]$:
\begin{eqnarray} \label{eq:e2rho}
\widetilde E[\rho]&=&E_0 +\mbox{tr}(h^0\delta{\rho})\\
&+& \frac{1}{2} \delta{\rho}_{ji,\sigma} \lbrack L_0 - S_0
^{\dagger} K_0^{-1} S_0\rbrack_{ijkl}^{\sigma\sigma'}
\delta{\rho}_{lk,\sigma'}.\nonumber
\end{eqnarray}
Notice that we could have derived Eq.~(\ref{eq:e2rho}) also
within the KR slave-boson approach. The corresponding
transformations for the derivatives are given in the Appendix.

The matrix $(L_0 - S_0 ^{\dagger} K_0^{-1} S_0) $ can be
considered as an effective interaction kernel between
particle-hole excitations in the GA. For the paramagnetic regime
this kernel reduces to the quasiparticle kernel of Vollhardt's
Fermi liquid analysis.~\cite{VOLLHARDT} Interestingly, the
off-diagonal elements of the matrices $K_{ij}$,
$L_{ijkl}^{\sigma\sigma'}$ and $S_{i,kl\sigma}$ can induce
intersite interactions between the GA quasiparticles. This is in
contrast with conventional HF theory of the Hubbard model which is
purely local.

The expansion of $E^{GA}_f[\rho,D]$, Eq.~(\ref{eq:fred}), is needed
up to first order only, since it is linear in the external field:
\begin{eqnarray}\label{eq:fdr}
E^{GA}_f[\rho,D]&=&F_0+\mbox{tr}(f^{GA}_0 \delta \rho) \nonumber\\
&+& \sum_{ijk\sigma\sigma'}  \rho^0_{ji,\sigma} f_{ij,\sigma}
\frac{\partial q_{ij\sigma}}{\partial \rho_{kk\sigma'}}
\delta \rho_{kk\sigma'} \nonumber\\
&+&  \sum_{ijk\sigma}
\rho^0_{ji,\sigma}f_{ij,\sigma}\frac{\partial
q_{ij\sigma}}{\partial D_{k}} \delta D_{k}.
\end{eqnarray}
Here $F_0=f^{GA}_0\rho^{(0)}$ describes the energy contribution
when the system would be frozen at the saddle-point level and we
used the fact that $q_{ij\sigma}$ does not depend on off-diagonal
densities. As before, the double occupancy fluctuations can be
eliminated through Eq.~(\ref{eq:ddr}). We define $\widetilde
E^{GA}_f[\rho]\equiv E^{GA}_f[\rho,D(\rho)]$ and
\begin{equation}
\widetilde  f_{ij\sigma}\equiv\frac{\partial \widetilde
E^{GA}_f[\rho]}{\partial\rho_{ji\sigma} }.
\end{equation}

In this paper we will restrict to density-density
and current-current  response functions with the current operator
given by:
\begin{equation}\label{CUOP}
{\bf J} =\sum_{\langle ij \rangle}{\bf j}_{ij},
\end{equation}
and
$${\bf j}_{ij} = - i \sum_{\sigma} t_{ij}
(c^{\dagger}_{i\sigma}c_{j\sigma}- c^{\dagger}_{j\sigma}c_{i\sigma}).$$
When only densities are involved  $f_{ij\sigma}$ is diagonal in the
site index, only $q_{ii\sigma}=1$
is present and the last two terms in Eq.~(\ref{eq:fdr}) vanish.
If currents are involved it is easy to show directly from
Eq.~(\ref{eq:fdr}) that the last two terms also vanish in the absence of
currents in the ground state, i.e.  $\widetilde  f_0= f^{GA}_0$.

Now, we proceed in analogy with the nuclear physics treatment of
effective mean-field theories in which the interaction potential
is density dependent.~\cite{RING,BLAIZOT} Indeed
Eq.~(\ref{eq:e2rho}) can be viewed as the energy expansion of an
effective  mean-field theory with the only difference that part of
the density dependence is due to the GA hopping renormalization
factors in the {\it kinetic} part of the Hamiltonian. The
advantage of this method with respect to other methods (e.g.,
equation of motion or diagrammatic methods) is that the present
derivation is solely based on the knowledge of an energy
functional associated with a Slater determinant which is precisely
what the Gutzwiller approximation provides.

The density matrix of an effective mean-field theory of this kind
obeys the equation of motion:~\cite{RING,BLAIZOT}
\begin{equation}
  \label{eq:motion}
  i\hbar\dot\rho=[\widetilde h[\rho]+ \widetilde  f (t),\rho],
\end{equation}
where we have defined an effective Gutzwiller Hamiltonian
\begin{equation}
  \label{eq:hgw2}
\widetilde  h_{ij\sigma}[\rho]=\frac{\partial\widetilde E }{\partial \rho_{ji\sigma}},
\end{equation}
which depends on densities only. At the saddle-point, we have
$\widetilde  h_0=\widetilde h[\rho^{(0)}]=h_0$.
The RPA is obtain by considering the limit of small amplitude
fluctuations in Eq.~(\ref{eq:motion}).

It is convenient to define the four sub-sectors of the fluctuations of
the density matrix using the projector properties of the density
matrix discussed above:
\begin{eqnarray}
\delta\rho^{hh}&\equiv&\rho^{(0)}\delta\rho\rho^{(0)}, \label{eq:fl1}\\
\delta\rho^{pp}&\equiv&\sigma^{(0)}\delta\rho\sigma^{(0)}, \label{eq:fl2}\\
\delta\rho^{hp}&\equiv&\rho^{(0)}\delta\rho\sigma^{(0)}, \label{eq:fl3}\\
\delta\rho^{ph}&\equiv&\sigma^{(0)}\delta\rho\rho^{(0)}. \label{eq:fl4}
\end{eqnarray}
The Slater determinant condition Eq.~(\ref{eq:sd}) implies that the
fluctuations, Eqs.~(\ref{eq:fl1})-(\ref{eq:fl4}), are not independent.
In fact, in terms of the fluctuations Eq.~(\ref{eq:sd}) reads:
\begin{equation}\label{SL1}
\delta\rho=\rho^{(0)}\delta\rho + \delta\rho\rho^{(0)} + (\delta\rho)^2.
\end{equation}
Projecting Eq.~(\ref{SL1}) onto the $hh$ and $pp$ sector of the saddle
point Slater determinant yields
\begin{eqnarray}
\delta\rho^{hh}&=&
-(1+\delta\rho^{hh})^{-1}\delta\rho^{hp}\delta\rho^{ph}\approx
-\delta\rho^{hp}\delta\rho^{ph}, \label{hhpp1}\\
\delta\rho^{pp}&=&
(1-\delta\rho^{pp})^{-1}\delta\rho^{ph}\delta\rho^{hp}\approx
\delta\rho^{ph}\delta\rho^{hp}, \label{hhpp2}
\end{eqnarray}
where the result on the right is valid in the small amplitude limit.
Thus it turns out that $pp$ and $hh$ density projections
are actually quadratic in
the $ph$ and $hp$ matrix elements. Therefore, on computing
$\widetilde h$ from
Eqs.~(\ref{eq:e2rho}) and~(\ref{eq:hgw2}) one should be aware of the
fact that the term
$\mbox{tr}(h^0\delta\rho)=
\sum_{\mu} \epsilon_{\mu} \rho_{\mu\mu}$ (which is first
order in the $pp$ and $hh$ density projections)
yields a quadratic contribution in the $ph$ and $hp$ matrix elements:
\begin{eqnarray}
\mbox{tr}(h^0\delta\rho)&=&\sum_{p} \epsilon_{p}\delta\rho_{pp}
                      +\sum_{h} \epsilon_{h}\delta\rho_{hh} \nonumber \\
&=& \sum_{ph}(\epsilon_p - \epsilon_h)\rho_{ph}\rho_{hp}.
\end{eqnarray}
In addition one can neglect the $pp$ and $hh$ matrix element in the last
term of Eq.~(\ref{eq:e2rho}).
Thus, up to second order in the particle-hole density fluctuations, one obtains
for the energy expansion Eq.~(\ref{eq:e2rho})
\begin{equation}\label{phe}
\widetilde E[\rho]=E_0 + \frac{1}{2}
(\delta\rho^{hp} ,\delta\rho^{ph})
  \left( \begin{array}{cc}
A & B \\
B^{*} & A^{*} \end{array}\right)
\left(\begin{array}{c}
 \delta\rho^{ph} \\
 \delta\rho^{hp}
\end{array}\right).
\end{equation}
Here the so called RPA matrices $A$ and $B$ are given by
\begin{eqnarray}
A_{ph,p'h'}&=& (\epsilon_p-\epsilon_h)\delta_{pp'}\delta_{hh'}
+ \frac{\partial\widetilde h_{ph}}{\partial \rho_{p'h'}}, \\
B_{ph,p'h'}&=&\frac{\partial\widetilde  h_{ph}}{\partial\rho_{h'p'}},
\end{eqnarray}
where the matrix $A$
contains matrix elements between particle-hole excitations whereas
the matrix $B$ is composed of matrix elements between the ground
state and two particle-hole excitations. $A$ and $B$ are related to
$M\equiv(L_0 - S_0 ^{\dagger} K_0^{-1} S_0)$ via
\begin{eqnarray*}
A_{ph,p'h'}&=& (\epsilon_p-\epsilon_h)\delta_{pp'}\delta_{hh'}\nonumber \\
&+& \sum_{ij\sigma,nm\sigma'}\!\!\psi_{i,\sigma}^{*}(p)\psi_{j,\sigma}(h)
M_{ij,nm}^{\sigma\sigma'}
\psi_{n,\sigma'}^{*}(h')\psi_{m,\sigma'}(p'), \nonumber \\
B_{ph,p'h'}&=&
\sum_{ij\sigma,nm\sigma'}\!\!\psi_{i,\sigma}^{*}(h)\psi_{j,\sigma}(p)
M_{ij,nm}^{\sigma\sigma'}
\psi_{n,\sigma'}^{*}(h')\psi_{m,\sigma'}(p'),
\end{eqnarray*}
and the transformation amplitudes $\psi_{i,\sigma}(\nu)$
have been defined in Eq.~(\ref{TRAFO1}).

To lowest order, we can now
linearize Eq.~(\ref{eq:motion}) retaining only  $ph$ and $hp$ matrix
elements:
\begin{equation}\label{eq:motionl}
  i\hbar\delta\dot\rho=[ h_0,\delta\rho]
+[\frac{\partial\widetilde h}{\partial\rho}\delta\rho+ \widetilde f,\rho^{(0)}],
\end{equation}
where we use the shorthand notation
\begin{equation}
\frac{\partial\widetilde h}{\partial\rho}\delta\rho=\sum_{ph}
\left(\frac{\partial\widetilde h}{\partial\rho_{hp}}\delta\rho_{hp}
+\frac{\partial\widetilde h}{\partial\rho_{ph}}\delta\rho_{ph}\right).
\end{equation}
Then from Eqs.~(\ref{eq:hgw2}),~(\ref{phe}) and~(\ref{eq:motionl})
one obtains the following linear response equation:
\begin{equation}
  \label{eq:rpa}
 \left\{ \left( \begin{array}{cc}
A & B \\
B^{*} & A^{*} \end{array} \right)-
\hbar \omega
\left(\begin{array}{cc}
1 & 0 \\
0 & -1
\end{array}\right)\right\}\left(\begin{array}{c}
\delta\rho^{ph}\\
\delta\rho^{hp}
\end{array}\right) =-
\left(
\begin{array}{c}
\widetilde f_{ph}\\
\widetilde f_{hp}
\end{array}\right) .
\end{equation}
This inhomogeneous equation can be solved by inverting the matrix on
the left-hand side which yields a linear relation between the external field
and the change in the density:
\begin{equation}\label{eq:lr}
\delta\rho=R(\omega)\widetilde f.
\end{equation}
We are now in the position to compute the
response of a one particle observable
$$O=
\sum_{ij\sigma} (o_{ij,\sigma}
c_{i\sigma}^{\dagger} c_{j\sigma} +h.c. ),$$
since in analogy with Eqs.~(\ref{eq:fdt}) and~(\ref{eq:fred}) its time
evolution is given by
\begin{equation}
\langle\Psi(t)|O|\Psi(t)\rangle=\sum_{ij\sigma}
 (o_{ij\sigma}^{GA}  \rho_{ji,\sigma}(t)  + h.c.),
\end{equation}
and the time evolution of $\rho$ is known from Eq.~(\ref{eq:lr}).

The linear response matrix $R(\omega)$ has poles at the
eigenfrequencies of the eigenvalue problem corresponding to
Eq.~(\ref{eq:rpa}) with $\widetilde f=0$:
\begin{equation}\label{eq:rpa2}
 \left\{ \left( \begin{array}{cc}
A & B \\
B^{*} & A^{*} \end{array} \right)
-
\hbar \Omega_{n}
\left(\begin{array}{cc}
1 & 0 \\
0 & -1
\end{array}\right)\right\}
\left(\begin{array}{c}
X^{(n)}\\
Y^{(n)}
\end{array}\right) = 0 .
\end{equation}
Here $\hbar \Omega_{n}\equiv E_n-E_0$  denote the excitation energies of
the system. In analogy with the HF+RPA approximation
the vacuum of these excitations is not the old starting GA state
 $|\Psi_0\rangle$ but a new state with both Gutzwiller type correlations
and RPA ground-state correlations.
We denote this state by $|\Phi_0\rangle$ and the corresponding
exited states by  $|\Phi_n\rangle$.

The matrix $R$ can be written in the following Lehmann
representation:
\begin{equation}\label{eq:leh}
R(\omega)_{ph,p'h'}=\sum_{n>0}\left\lbrack
\frac{X_{ph}^n X_{p'h'}^{n*}}{\omega-\Omega_{n}+i\epsilon} -
\frac{Y_{p'h'}^n Y_{ph}^{n*}}{\omega+\Omega_{n}+i\epsilon}\right\rbrack.
\end{equation}
In analogy with the HF+RPA method, we introduce the
following notations:
\begin{eqnarray}
\langle 0 | a^{\dagger}_h a_p |n \rangle&\equiv& X_{ph}^{n}, \label{eq:xy1}\\
\langle 0 | a^{\dagger}_p a_h |n \rangle&\equiv&Y_{hp}^{n}. \label{eq:xy2}
\end{eqnarray}
The states  $|n\rangle$ are not true excitations
of the system but represent auxiliary notational objects. Roughly
speaking, they can be thought of as RPA states without the
Gutzwiller projector. For example  $|0\rangle$ is the analog
of the state $|SD\rangle$ but at RPA level (it contains RPA ground-state
correlations but lacks Gutzwiller correlations). We will call them
unprojected RPA states. The eigenvector
$(X_{ph}^{(n)},Y_{hp}^{(n)})$ can be identified with the
particle-hole and hole-particle components of the unprojected RPA
excited state $|n\rangle$ with respect to the unprojected RPA
ground state $|0\rangle$.

Schematically the four states are related in the following
way:
\begin{eqnarray}
|SD \rangle &{P \atop \longrightarrow}&|\Psi_0\rangle\nonumber\\
 RPA \downarrow & &\downarrow RPA\nonumber\\
|0\rangle  &{P \atop \longrightarrow}&|\Phi_0\rangle, \nonumber
\end{eqnarray}
where $P$ indicates Gutzwiller projection.

Within the above formalism, it is straightforward to evaluate the
current-current correlation function. The real part of the optical
conductivity consists of a Drude part at $\omega=0$ and a regular
part for $\omega>0$:
\begin{equation}
\sigma(\omega)=D\delta(\omega)+ \pi \sum_{n>0} \frac{|\langle
\Phi_n
|j_{\alpha}|\Phi_0\rangle|^2}{E_n-E_0}\delta(\omega-(E_n-E_0)).
\end{equation}
With the above approximations and notations
$$
\langle \Phi_n|j_{\alpha}|\Phi_0\rangle=\langle
n|j^{GA}_{\alpha}|0\rangle,$$
where the matrix element on the right can be evaluated using
Eqs.~(\ref{TRAFO1}),~(\ref{eq:gaop}),~(\ref{eq:xy1}), and~(\ref{eq:xy2}).
Obviously the
matrix elements within $\sigma(\omega)$ are renormalized by the GA
hopping factors whereas $R(\omega)$ does not contain such
renormalization. Thus the latter quantity does not correspond to a
physical response function within the GA+RPA approach.

The Drude weight $D$ can be obtained from the f-sum rule (see $S^{-1}$
in next section)
\begin{equation}\label{fsum}
\int_0^{\infty}\,d\omega \sigma(\omega) = -\frac{1}{2}\pi
\langle T_{\alpha} \rangle_{GA},
\end{equation}
where the kinetic energy in $\alpha$-direction $\langle
T_{\alpha} \rangle_{GA}$ is evaluated within the GA.

In practice, for computational purposes one can use all the
standard formulas of the HF+RPA scheme by  substituting true
operators with GA effective operators and excitations by the
unprojected excitations $|n\rangle$.

The matrix elements
$\langle \Phi_0|O|\Phi_n\rangle=\langle 0|O^{GA}|n\rangle$ can be used to
characterize a given RPA excitation. Specific examples which will
be considered below are the transition density
\begin{equation}\label{eq:trdens}
\delta n_i^{m}\equiv \langle 0|\hat n_i|m \rangle,
\end{equation}
and the transition current
\begin{equation}\label{eq:trcurr}
\delta {\bf j}_{ij}^{n} \equiv \langle 0|{\bf j}_{ij}^{GA}|n\rangle,
\end{equation}
which can be  interpreted as follows. Consider a wave packet
$$|\psi(t) \rangle=
\exp(-i E_0 t) |\Phi_0\rangle+\eta \exp(-i E_m t)|\Phi_m\rangle,$$
consisting of a small admixture $\eta$ of an exited state $m$ to
the ground state. For example this can be the result of an
excitation of the mode $m$ by an appropriate weak external
perturbation.  The time dependent expectation value of the charge
is then given by:
$$
\langle \psi(t)|\hat n_i|\psi(t) \rangle=\langle 0|\hat n_i|
0\rangle+ \eta \delta n_i^{m} e^{-i \Omega_m t} + h.c.,$$ and an
analogous expression holds also for the current. Here $\langle
0|\hat n_i| 0\rangle=\langle SD|\hat n_i| SD\rangle$ since
one-particle densities are not renormalized by the RPA. We see
that the transition charges and currents  are proportional to the
amplitude of the time dependent  fluctuation  that would occur at
frequency $\Omega_{m}$ if the state $m$ is excited by a weak
perturbation~\cite{RING} (see also Ref.~\onlinecite{LOR2}).

\subsection{Sum rules}
Sum rules form a very important tool in the theory of collective
excitations. In many cases they allow us to calculate global
properties in a simple way and therefore they are useful in
testing different approximation schemes. In general, a sum rule is
related to the k-th moment of the excitation strength distribution
produced by a single-particle operator $O$ (see e.g.,
Ref.~\onlinecite{MAHAN}):
\begin{equation}
S^k\equiv \sum_{n}(E_n-E_0)^k
|\langle\Psi_n|O|\Psi_0\rangle|^2.
\end{equation}
Within the present scheme we have:
\begin{equation}
S^k=\sum_{n}(E_n-E_0)^k
|\langle n|O^{GA}| 0\rangle|^2,
\end{equation}
and we restrict to current or density operators for $O$.
The energy sum rule $S^1$  can be written as a double commutator
\begin{equation}\label{SR1}
S^1 = \sum_{n=1}(E_n-E_0) |\langle\Psi_n|O|\Psi_0\rangle|^2
=\frac12 \langle \Psi_0 | [O, [H,O]|\Psi_0\rangle.
\end{equation}
In analogy with the derivation by Thouless~\cite{THOULESS}, one
can show that the sum rule Eq.~(\ref{SR1}) is satisfied if the
left-hand side is evaluated at the GA+RPA level  and the
right-hand side is calculated using the GA ground state wave
function. The same applies for the $S^{-1}$ sum rule, which in
case of the optical conductivity discussed above corresponds to
the f-sum rule.

In the following, we consider as an example the first
moment of the charge-charge correlation function by setting $O
\equiv n_i=\sum_{\sigma}n_{i\sigma}$ for some lattice site ${\bf R}_i$.
It is straightforward to
evaluate the double commutator and we find for the corresponding
sum rule
\begin{equation}\label{SR2}
S^1_n= - 2 \sum_{m,i} (E_m-E_0) |\langle
0|n_i|m \rangle|^2= 2\langle T \rangle_{GA},
\end{equation}
where $\langle T \rangle_{GA}$ denotes the kinetic energy
evaluated within the GA.

\begin{figure}[tbp]
\includegraphics[width=8cm,clip=true]{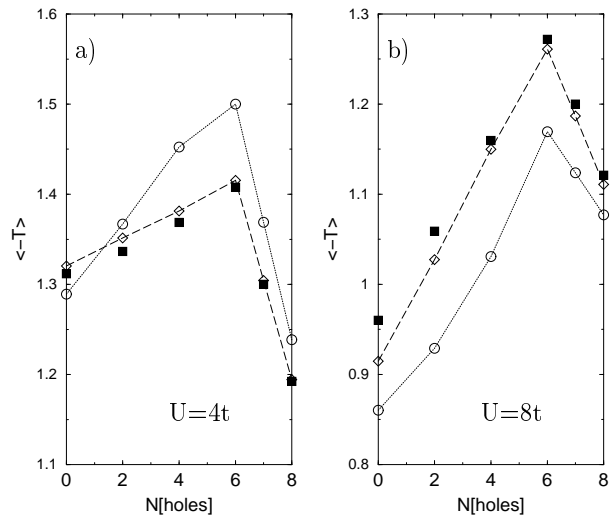}
\caption{\baselineskip .185in
Kinetic energy per site of a $4\times 4$ Hubbard
model for $U/t=4$ (a) and $U/t=8$ (b) with periodic boundary conditions
as a function of hole doping. Filled squares: exact result;
Circles: Unrestricted HF approximation; Diamonds: Unrestricted GA.}
\label{fig:1}
\end{figure}

The sum rules of Eqs.~(\ref{fsum}) and~(\ref{SR2}) provide a first
encouraging argument that the unrestricted GA could improve the
description of charge fluctuation with respect to the
corresponding HF method. This is based on the fact that the GA
kinetic energy is already renormalized on the mean-field level. In
Fig.~\ref{fig:1} we compare the exact kinetic energy with
unrestricted GA and HF results for various hole concentrations in
a Hubbard model ($4\times 4$ lattice) with nearest neighbor
hopping $t_{ij}=-t$. The GA Slater determinants have been obtained
using the method described in Ref.~\onlinecite{GOETZ1}. Notice
that for this small system it is in general not a problem to find
the true mean-field ground state via the variational procedure. We
usually performed several runs starting from different initial
configurations and checked the stability of the resulting states
by adding some noise to the solutions. These are in general
characterized by an inhomogeneous charge distribution except for
the closed shell configurations.

For small values of $U/t$ there is almost perfect agreement between
the GA method and exact results. In this limit, where kinetic
effects dominate the correlation part, HF overestimates the value
of $\langle-T\rangle$ since the corresponding quasiparticle hopping between
sites $i$ and $j$ is described by the bare matrix element t$_{ij}$. On
the other hand, for $U/t=8$ the large HF onsite renormalization
(corresponding to an overestimate of the spin-polarization) is the
origin for the kinetic energy to be lower than the exact result.
In contrast, the values of $\langle-T\rangle$ in the GA approximation
correctly reproduce the exact result, especially in high-doping
regime, where the spin density is reduced in large parts of
the lattice. It follows that the first moment of a density-density
correlation function will be more accurate in GA+RPA than in the HF+RPA
approximation. The same holds for the $S^{-1}$-sum rule for
the optical conductivity.

To summarize this section, the idea of our method is to supplement
the Gutzwiller approximation with RPA fluctuations analogous to
the HF+RPA approach.~\cite{HFRPA} Since the GA provides a much
better initial saddle-point than HF one can expect that the
fluctuation corrections within GA+RPA will allow us for a more
accurate description of correlation effects than the HF+RPA
approach. In the remaining sections we analyze in detail small
data cases  to test the domain of applicability of the method and
finally show some applications in larger systems.

\section{Two-site Hubbard model}
\label{two}
In order to demonstrate the method developed
above, we consider in the following the two-site Hubbard model which can be
solved exactly and can be studied analytically in both the
GA+RPA and  HF+RPA approximations.

Exact ground-state energy and double occupancy at
half filling (i.e., two particles) are given by
$$E_0=\frac{U}{2}(1-\sqrt{1+(4t/U)^2}),$$
and
$$\langle n_{\uparrow}n_{\downarrow}\rangle=
\frac{1}{2}\frac{E_0^2}{4t^2 + E_0^2},$$
respectively.
The exact optical conductivity displays one transition between
the ground state and an excited
state with energy $E_U=U$ resulting in  the excitation energy
$$E_U-E_0=\frac{U}2 (1+\sqrt{1+(4t/U)^2}).$$
The corresponding matrix element of the current operator is
$$|\langle 0|j|U \rangle|^2=\frac{16t^4}{(E_0^2+4t^2)}.$$
Upon minimizating the GA energy functional of the half-filled
two-site model one finds a paramagnetic solution below
$$U_{crit}^{GA}/t =  8(\sqrt{2}-1)\approx 3.31,$$
and a Ne\'el-type state with $m_1^0=-m_2^0$  where
$$m_i=\langle n_{i,\uparrow}\rangle - \langle
n_{i,\downarrow}\rangle \neq 0,$$
for $U > U_{crit}^{GA}$.
Within HF theory the corresponding critical value is
$U_{crit}^{HF}/t=2$.

Clearly the transition in either case is non-physical since it does not
occur in the exact solution. In this sense
the increase of  $U_{crit}^{GA}$ with respect to $U_{crit}^{HF}$ is in favor
of the GA since it extends the parameter range of the right singlet
paramagnetic solution. At large $U$, disregarding the non-physical broken
symmetry, the Ne\'el-type state in GA
allows the system to reduce the double occupancy and at the same time
prevents the occurrence of the
Brinkmann-Rice (BR) transition towards localization at $U_{BR}=8t$.

Since the analytic expressions for the symmetry-broken regime
become quite lengthy
we restrict the derivation below to the paramagnetic case. In this
limit the mean-field part of the energy  is given by
\begin{equation}
\mbox{tr}(h_0 \rho)=t(1-u^2)\sum_{\sigma} (\rho_{pp,\sigma}-\rho_{hh,\sigma}),
\end{equation}
which defines the
diagonal Gutzwiller Hamiltonian in Eq.~(\ref{eq:hgw}). The hole
(particle) state is the bonding (antibonding) state and $u=U/(8t)$.

The GA kinetic energy reads as
$$E_{kin}=-2t(1-u^2),$$
and the expansion of the GA energy functional leads to
[see Eqs.~(\ref{eq:lks})]
\begin{eqnarray}\label{eq:lks2}
L_{ii,ii}^{\sigma\sigma}&=&\frac{4 t \left( 2 + 3 u^2 - u^4 \right) }{1 - u^2}, \\
L_{ii,ii}^{\sigma,-\sigma}&=&\frac{8 t}{1 - u}, \\
L_{ii,jj}^{\sigma\sigma'}&=&-\frac{8 t u^2}{1 - u^2}\ \ \ \ i\ne j, \\
L_{ii,ij}^{\sigma\sigma'}&=&2tu\ \ \ \ i\ne j, \\
K_{ii}&=&\frac{32 t}{1 - u^2}, \\
K_{ij}&=&-\frac{32 t u^2}{1 - u^2}\ \ \ \ i\ne j, \\
S_{i,ii\sigma}&=&-\frac{16 t}{1 - u^2}, \\
S_{i,jj\sigma}&=&\frac{16 t u^2}{1 - u^2}\ \ \ \ i\ne j, \\
S_{i,ij\sigma}&=&-4 t u \ \ \ \ i\ne j.
\end{eqnarray}
One of the peculiarities of the present approach is the appearance of
onsite interactions for quasiparticles with the same
spin ($L_{ii,ii}^{\sigma\sigma}\neq 0$).
These interactions do not occur in the standard RPA since they would
violate the Pauli exclusion principle but appear here because
of the density dependence of the effective interaction between
particles. Notice also that many of the matrix elements would diverge
at the BR transition if it were not hidden by the
spin-density wave (SDW) transition.

Eliminating the double occupancy fluctuations with help of the antiadiabatic
condition Eq.~(\ref{AAC}), the following interaction matrix is obtained:
\begin{eqnarray}\label{eq:m2}
M_{ii,ii}^{\sigma\sigma}&=&\frac{4 t u^2 \left( 3-u^2\right) }{1 - u^2}, \\
M_{ii,ii}^{\sigma,-\sigma}&=&\frac{ 8 t u}{1 - u^2}, \\
M_{ii,jj}^{\sigma\sigma'}&=&0\ \ \ \ i\ne j, \\
M_{ii,ij}^{\sigma\sigma'}&=&0\ \ \ \ i\ne j, \\
M_{ij,ij}^{\sigma\sigma'}&=&M_{ij,ji}^{\sigma\sigma'}=-t u^2\ \ \ \ i\ne j.
\end{eqnarray}
Remarkably, intersite interactions vanish except for the appearance of a new
interaction term between off-diagonal charges
($M_{ij,ij}^{\sigma\sigma'}$). Using
Eqs.~(\ref{hhpp1}) and~(\ref{hhpp2}) one can show that these new
off-diagonal interactions do not contribute to the RPA matrices and the
expansion of the energy reads as:
\begin{equation}
E=E_{GA} + \sum_{i=1}^2 \lbrack U^s (\delta m_i)^2 + U^c (\delta \rho_i)^2
\rbrack,
\end{equation}
where the charge- and spin-interaction coefficients are given by
\begin{eqnarray}
U^c &=& u \frac{(2-u)(1+u)}{1-u} t, \\
U^s &=& -u \frac{(2+u)(1-u)}{1+u} t.
\end{eqnarray}
and naturally coincide with the Landau parameters $F_0^s$, $F_0^a$
derived by Vollhardt in Ref.~\onlinecite{VOLLHARDT}.

The RPA-matrices read as
\begin{eqnarray}
A&=&\left( \begin{array}{cc} \Delta E + U^+ & U^- \\ U^- & \Delta E + U^+
\end{array} \right), \\
B&=&\left( \begin{array}{cc} U^+ & U^- \\ U^- & U^+
\end{array} \right),
\end{eqnarray}
with the GA particle-hole excitation energy $\Delta E = 2t(1-u^2)$
and $U^{\pm}=U^c \pm U^s$.

Upon diagonalizing the RPA problem one obtains the
eigenvalues:
\begin{equation}
\omega_{\pm}^2=\Delta E \lbrack \Delta E + 2(U^+ \pm U^-) \rbrack,
\end{equation}
which correspond to a singlet ($\omega_{+}$) and a magnetic
excitation ($\omega_{-}$), respectively. The former is the charge excitation
which contributes to the optical conductivity whereas the latter
can be identified with the Goldstone mode driving the
transition from the paramagnetic
to the SDW state. This transition occurs at
$\omega_{-}^2=0=\Delta E+4U^s$ so that
the transition to the symmetry-broken state
is only determined by the spin interaction $U^s$.
Notice that for the HF approximation we have $U^c=U/4=-U^s$
so that in this case the transition occurs at $U_{crit}^{HF}/t=2$ as stated
above.

When we expand the RPA charge excitation energy of the
GA approach for small $U/t$ we obtain
\begin{eqnarray}
\omega_{+}^{2}&=&2t(2t+U)+\frac14 U^2 +{\cal O}(U^4) \ \ {\rm (GW+RPA)}, \\
\omega_{+}^{2}&=&2t(2t+U) \ \ \ \ \ \ \ \ \ \ \ \ \ \  \ {\rm (HF+RPA)},
\end{eqnarray}
which has to be compared with the expansion of the exact excitation
\begin{equation}
\omega_{exact}^2=2t(2t+U)+\frac12 U^2 +{\cal O}(U^3),
\end{equation}
Thus the RPA corrections to the Gutzwiller approximation partially includes
higher order contributions in U which are not contained in the
HF+RPA approach.

The eigenvectors of Eq.~(\ref{eq:rpa2}) are given by
\begin{eqnarray}
X^{\pm}_{\uparrow} &=& 1/2 \alpha_{\pm} (1+\omega_{\pm}/(\Delta E)), \\
X^{\pm}_{\downarrow} &=& \pm 1/2 \alpha_{\pm} (1+\omega_{\pm}/(\Delta E)), \\
Y^{\pm}_{\uparrow} &=& 1/2 \alpha_{\pm} (1-\omega_{\pm}/(\Delta E)), \\
Y^{\pm}_{\downarrow} &=& \pm 1/2 \alpha_{\pm} (1-\omega_{\pm}/(\Delta E)),
\end{eqnarray}
and the normalization factor is
$\alpha_{\pm}^2=\Delta E /(2 \omega_{\pm})$.

\begin{figure}[tbp]
\includegraphics[width=8cm,clip=true]{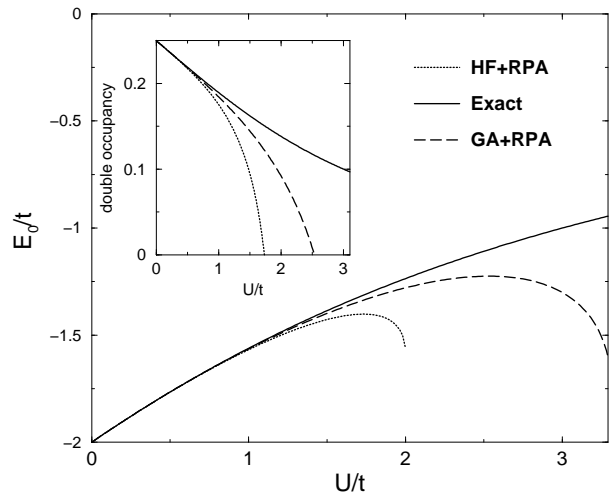}
\caption{Comparison of the GA+RPA, HF+RPA and exact results for the
ground-state energy $E_0$ of the two-site Hubbard model.
The inset shows the double occupancy as a function of $U/t$ within the
same approaches.}
\label{fig:2}
\end{figure}

We are now able to compute the RPA double occupancy
by evaluating the corresponding correlation function as
$\sum_{m=\pm}\langle 0|n_{\uparrow}|m\rangle\langle m|n_{\downarrow}
|0\rangle$ leading to
\begin{equation}
D^{RPA}_{GA}=\frac14+\frac18 \Delta E\left( 1/\omega_{+}-1/\omega_{-}\right).
\end{equation}
Approaching the SDW transition leads to an increase of spin
fluctuations which leads to a divergent D$^{RPA}$ at the
transition point due to the Goldstone mode $\omega_{-}\rightarrow
0$ (see inset of Fig.~\ref{fig:2}). From the double occupancy we
can compute the corrections to the ground state energy using the
coupling constant integration trick~\cite{MAHAN,FETTER}, i.e.
$E_{int}=\int_0^{U}dx D^{RPA}(x)$ which yields
\begin{eqnarray}
E_{int}^{GA+RPA}&=&-2t+U/2 +2t \left(-2+\right.\nonumber \\
&+& \left.\sqrt{2-(1-u)^2}+\sqrt{2-(1+u)^2}\right).
\end{eqnarray}
We thus obtain the GA+RPA ground-state energy
which is displayed in Fig.~\ref{fig:2} together with the corresponding
HF+RPA and the exact result.

Naturally for such a small system any mean-field treatment will
result in large errors, however, due to the increased value of
$U_{crit}^{GA}$ and the extra contributions in $U$ discussed above
GA+RPA performs much better than HF+RPA.
In general, RPA-corrections overshoot the exact energy
which becomes significant when one approaches the non-physical SDW
transition.

For larger systems where the broken symmetry mean-field can
correspond to long-range order or quasi long range order in the
exact solution the agreement is much better as shown in
Ref.~\onlinecite{GOETZ3}. In addition the performance improves
with increasing dimensionality as in any mean-field+RPA theory.

Finally it is straightforward to check that standard
sum rules are obeyed. For example the current operator matrix elements
between ground state and the charge ($+$) and the
magnetic ($-$) excitation are given by:
\begin{eqnarray}
|\langle 0|j^{GA}|-\rangle|^2 &=& 0, \\
|\langle 0|j^{GA}|+\rangle|^2 &=& t(1-u^2) \omega_{+},
\end{eqnarray}
as a result one thus obtains that the f-sum rule
$|\langle 0|j^{GA}|+\rangle|^2/\omega_{+} + (1/2) E_{kin}^{GA} =0 $
is satisfied within the GA+RPA approach.

\section{Response functions in the 1D and 2D  HUBBARD model}
\label{dynamics}

In this section we apply our method to the calculation of response
functions in the 1D and 2D Hubbard model. In the first part, we
show that already on the saddle-point level, the GA yields rather
accurate excitation energies as compared to HF. Inclusion of RPA
corrections then lead to an additional redistribution of spectral
weight in the correlation functions, which is demonstrated in the
second part by a detailed comparison of exact diagonalization
results with GA+RPA and HF+RPA. Especially this
comparative study is intended for an a posteriori justification
of the antiadiabaticity assumption Eq.~(\ref{AAC}).

\subsection{Optical conductivity of the half-filled Hubbard model in the
Gutzwiller approximation} \label{sec:saddle} After
Ref.~\onlinecite{GOETZ3} was published we become aware that the
RPA residual interaction for the SDW vanishes in the channel
relevant for the optical conductivity (zero momentum and odd
parity). This is a pathology of the Hubbard model at half-filling
and occurs both in the HF+RPA and in the GA+RPA, whereas it does
not occur for more complicated models (like multiband Hubbard) or
symmetry-broken ground-states, like polarons or stripes. As a
consequence the optical conductivity $\sigma(\omega)$ is the same
on the GA+RPA and GA level. It is nevertheless quite instructive
to examine $\sigma(\omega)$ within the mean-field approximation
since this demonstrates the better starting saddle-point of GA in
comparison to HF, with respect to the one-particle excitation
energies. For this purpose, we compare in the following the GA and
HF optical conductivity with numerical results and present the
results for charge-charge correlation functions where the
mentioned pathology does not occur and instead GA+RPA introduces
non trivial corrections to the dynamical response functions.

\begin{figure}[tbp]
\includegraphics[width=8cm,clip=true]{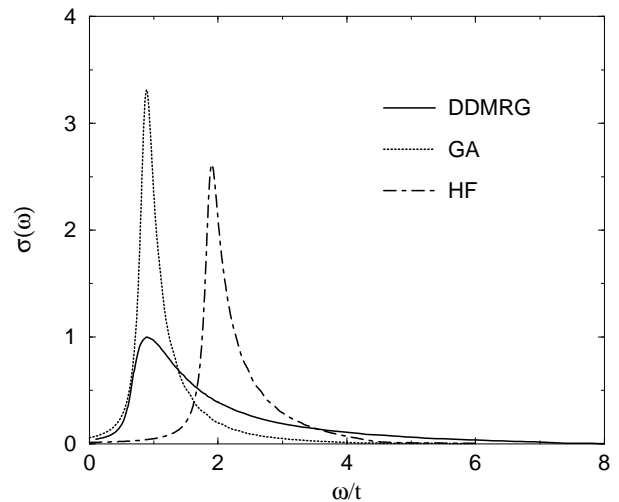}
\caption{Optical conductivity of the 1D Hubbard chain (120 sites) for $U/t=3$.
Solid: DDMRG; Dotted: GA; Dashed-dotted: HF. DDMRG data by courtesy of
E. Jeckelmann.
$\sigma(\omega)$ has been convoluted with a Lorentzian with width
$\varepsilon=0.1t$.}
\label{fig:oc1}
\end{figure}

Fig.~\ref{fig:oc1} displays $\sigma(\omega)$ for a half-filled Hubbard
chain and $U/t=3$ convoluted with a Lorentzian
$L(\omega)=\varepsilon/[\pi(\varepsilon^2+\omega^2)]$ and $\varepsilon=0.1t$.
For both GA and HF, the ground state is characterized
by long-range  SDW order and the regular part of $\sigma(\omega)$ is given by
\begin{equation}
\label{eq:sigmamf}
\sigma(\omega)=\frac{\Delta^2}{2\omega^2}\Re \sqrt{\frac{16t_{eff}^2}
{\omega^2-\Delta^2}-1},
\end{equation}
where $t_{eff}$ contains the $z$-factor renormalization in case of
the GA ($t_{eff}^{GA}=tz_\uparrow z_\downarrow$), $t_{eff}^{HF}=t$
and  $\Re$ denotes the real part. The SDW gap in HF is related to
the onsite magnetization $\Delta^{HF}=2 U |S_z|$, whereas within
the Kotliar-Ruckenstein  formulation of the GA \cite{KR} it is
determined by the difference in the local spin-dependent Lagrange
multipliers $\Delta^{GA}=\lambda_{\uparrow}-\lambda_{\downarrow}$.
It is quite interesting to observe that the onset of excitations
coincides rather well in the DDMRG~\cite{JECKEL} and GA
approaches, whereas the HF gives a gap which is by far to large.
However, although GA leads to excellent results for the gap
energies it turns out that the corresponding intensity is
overestimated. This has the consequence that most of the
high-frequency evolution of $\sigma(\omega)$ is compressed close
to the threshold, whereas the DDMRG shows a much broader spectrum.
We have checked the broadness of the exact spectrum by performing
exact diagonalization in small clusters. In fact from
Eq.~(\ref{eq:sigmamf}) one obtains that the large frequency tail
for GA and HF behaves as $\sigma(\omega\gg\Delta)\sim 1/\omega^3$
whereas the correct field theoretical  result is
$\sigma(\omega\gg\Delta)\sim 1/\omega$.~\cite{JECKEL}

To summarize, the optical conductivity results are the same at
mean-field or mean-field plus RPA, and, therefore, no corrections are
introduced by our method in this (rather pathological) case. As
compared to the DDMRG results, GA performs much better than HF since
it reproduces the onset of excitations with a better accuracy.
However, the width of the spectrum is underestimated in both
mean-field approaches.

\subsection{Comparison with exact results} \label{sec:exa}

In order to compare the GA+RPA approach with exact results, we
investigate the onsite density-density response function

\begin{equation}
S_c(\omega)=\sum_i\sum_{m>0}|\langle \Psi_m
|n_i|\Psi_0\rangle|^2\delta(\omega-(E_m-E_0)).
\end{equation}

In this case GA and GA+RPA solutions are different since the
pathology discussed in the previous subsection is not present.

The following sum rule is obeyed [see Eq.~(\ref{SR2})]:

\begin{equation}\label{SR3}
\int\!\mbox{d}\omega \omega S_c(\omega) = -\langle T \rangle_{GA}.
\end{equation}

\begin{figure}[tbp]
\hspace*{-0.2cm}\includegraphics[width=7.5cm,clip=true]{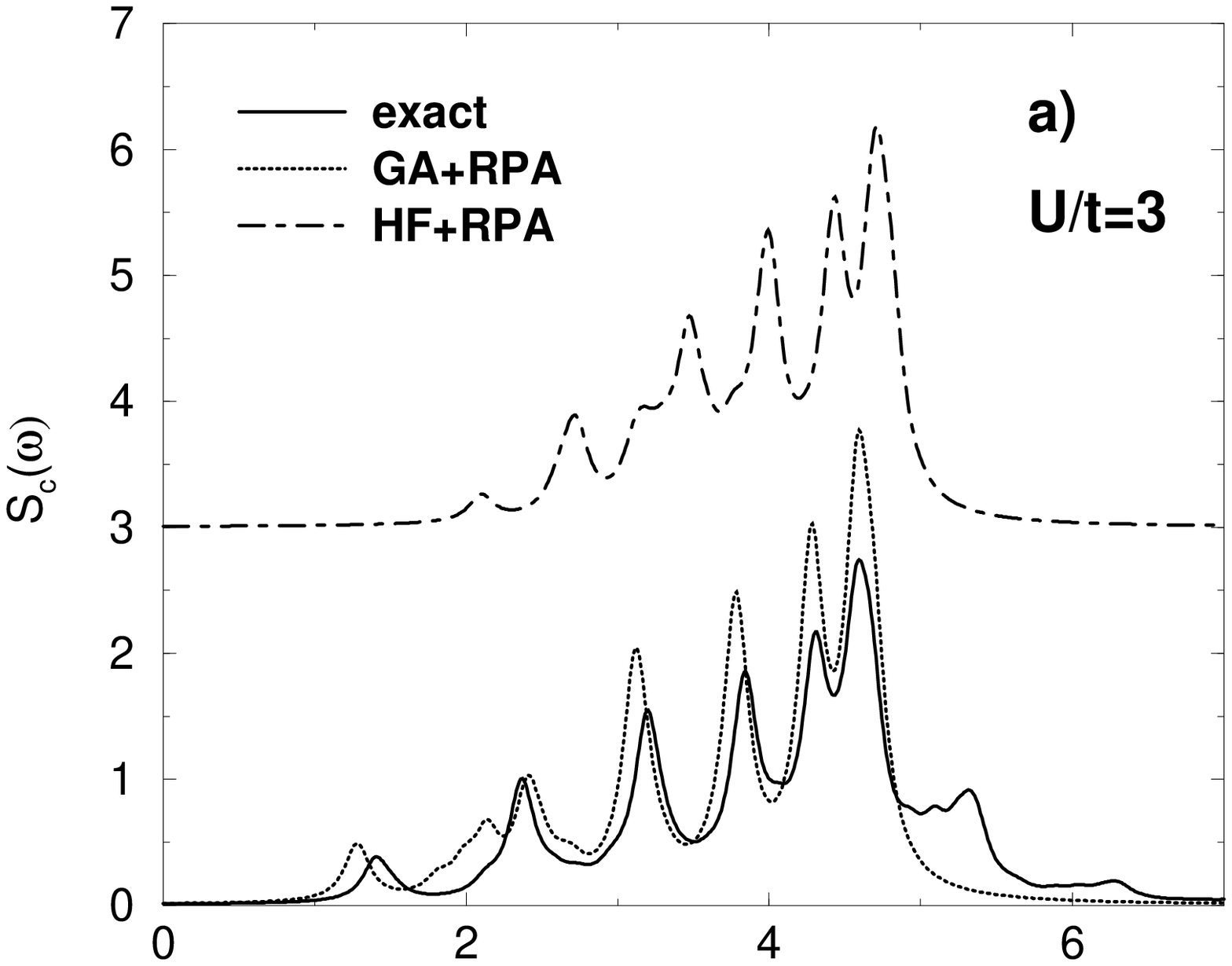}
\includegraphics[width=7.5cm,clip=true]{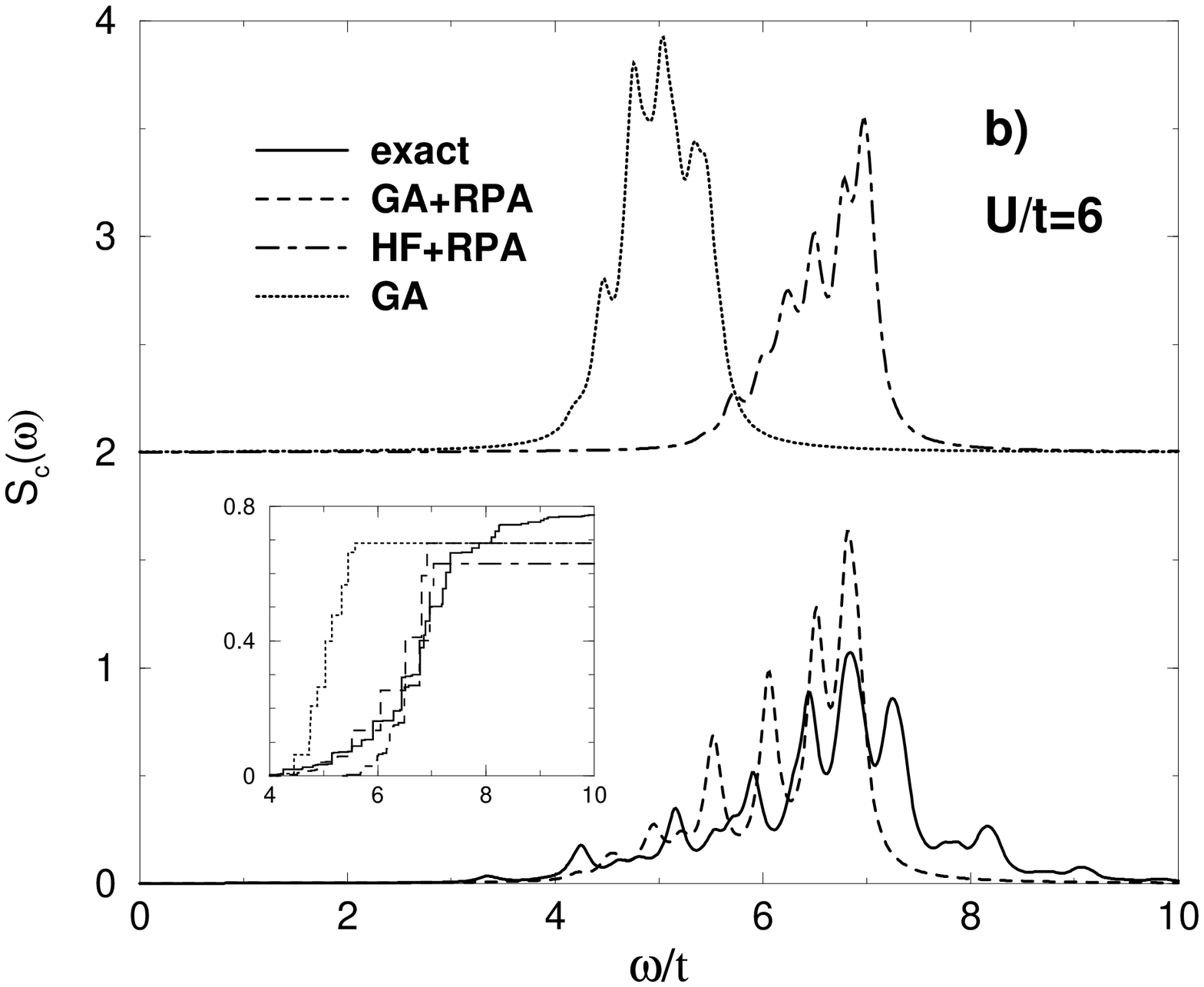}
\caption{Charge correlation function $S_c(\omega)$ for a
half-filled Hubbard chain (14 site)
in case of $U/t=3$ (a) and $U/t=6$ (b). For clarity the curves for
HF+RPA and GA (in b) have been shifted upwards. The inset in (b) shows the
integrated weight as a function of frequency for GA and GA+RPA, respectively.}
\label{fig:n14}
\end{figure}

In Fig.~\ref{fig:n14} we show $S_{c}(\omega)$ for a half-filled
Hubbard chain with 14 sites calculated with exact
diagonalization, GA+RPA, and HF+RPA. For $U/t=3$ the
lowest energy excitation is at $\omega \sim 1.4t$ (exact),
$\omega \sim 1.3t$ (GA+RPA) and
$\omega \sim 2.1t$ (HF+RPA), respectively, so that GA+RPA is much more accurate
than the standard HF+RPA approach. Also the higher energy
excitations computed with GA+RPA are in remarkable
agreement with the exact results. Moreover, the oscillator strength
of the two lowest excitations coincides rather well with the
intensity obtained with exact diagonalization. The small high-energy
features between $\omega \sim 5t$ and $\omega \sim 7t$ present in the exact
result do not show up in the GA+RPA correlations. As a consequence,
GA+RPA slightly overestimates the intensity between
$\omega \sim 3t$ and $\omega \sim 5t$,
since  from $\langle T\rangle_{GA} \approx \langle
T\rangle_{exact}$, the sum rule requires the integrated spectral
weight to be approximately the same in both GA and the exact
result.

The accuracy of the GA+RPA approach is also remarkable
with respect to the fact that the underlying mean-field solution
(GA) corresponds to a SDW state, whereas the exact solution in 1D
does not show long-range order. However, it is well known that
correlation functions decay slowly and thus are quasi-long ranged.
Hence, only for very low energies one expects disagreement due to
this problem and since excitations in the system under
consideration are gapped this discrepancy does not really show up.

Fig.~\ref{fig:n14}(b) reveals that GA+RPA provides a better description of
the low-energy excitations than HF+RPA, even at larger values of $U/t$.
In addition, we show in Fig.~\ref{fig:n14}(b)
the charge-charge correlations for the bare GA. In this case,
the corresponding
excitations are located in a narrow energy window on the low-energy side
of the exact spectrum with rather incorrect oscillator strength.
Thus RPA corrections induce the broadening and shift
of excitations to higher energies with a simultaneous redistribution of
intensity. This can also be deduced from the inset to Fig.~\ref{fig:n14}(b)
which shows the evolution of the integrated spectrum $\int_0^{\omega}
S_c(\nu)d\nu$ for GA and GA+RPA.
From the sum rule Eq.~(\ref{SR3}) it is obvious that both GA and GA+RPA
approach the same integrated spectral intensity but with the GA+RPA
spectral
evolution broadened and shifted to higher energies with respect to the
bare GA.

\begin{figure}[tbp]
\includegraphics[width=7.5cm,clip=true]{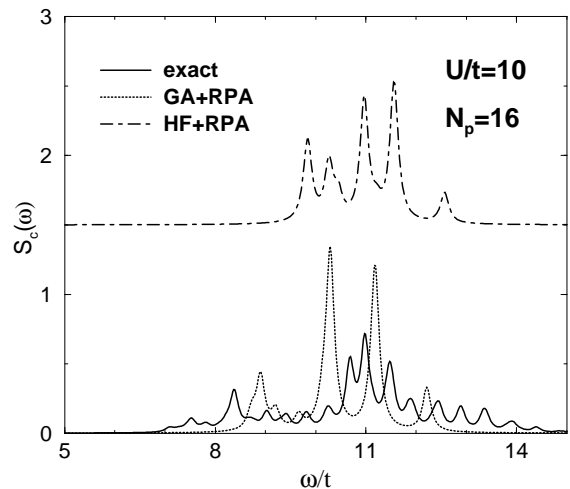}
\caption{Charge correlation function $S_c(\omega)$ for the
half-filled $4\times 4$ Hubbard model (16 particles) 
in case of $U/t=10$.}
\label{fig:44u10}
\end{figure}

In systems with large dimensionality we expect our approach to
improve, since the GA is an exact solution of the Gutzwiller
variational problem in infinite dimensions. Furthermore, quite
generally mean-field theories become better as the dimensionality
is increased.
Fig. ~\ref{fig:44u10} displays S$_c(\omega)$ for a half-filled 
$4\times 4$ system (i.e. 16 particles) with $U/t=10$.
The lowest prominent energy peak in the  exact solution 
occurs at $\omega_{min}^{ex}=8.4t$ and a second bunch of excitations starts at 
$U/t \approx 10t \dots 11t$ 
decaying in intensity towards higher energy.
The lowest energy excitation within 
GA+RPA ($\omega_{min}^{GA+RPA}=8.7t$)
is remarkably close to the exact value in contrast to HF+RPA where the
lowest peak appears at $\omega_{min}^{HF}=9.8t$.
Moreover, the center of high energy excitations in the exact solution 
is represented by two peaks in the GA+RPA spectrum at
$9.7t$ and $11.2t$ whereas they are shifted to slightly higher energy 
within HF+RPA. 
It is interesting to observe that
also in this energy range the GA+RPA method gives a better (although
rather crude) approximation than HF+RPA despite the expected 
failure of the antiadiabaticity condition Eq.~(\ref{AAC})
for energies larger than the Mott-Hubbard gap.

\begin{figure}[tbp]
\includegraphics[width=7.5cm,clip=true]{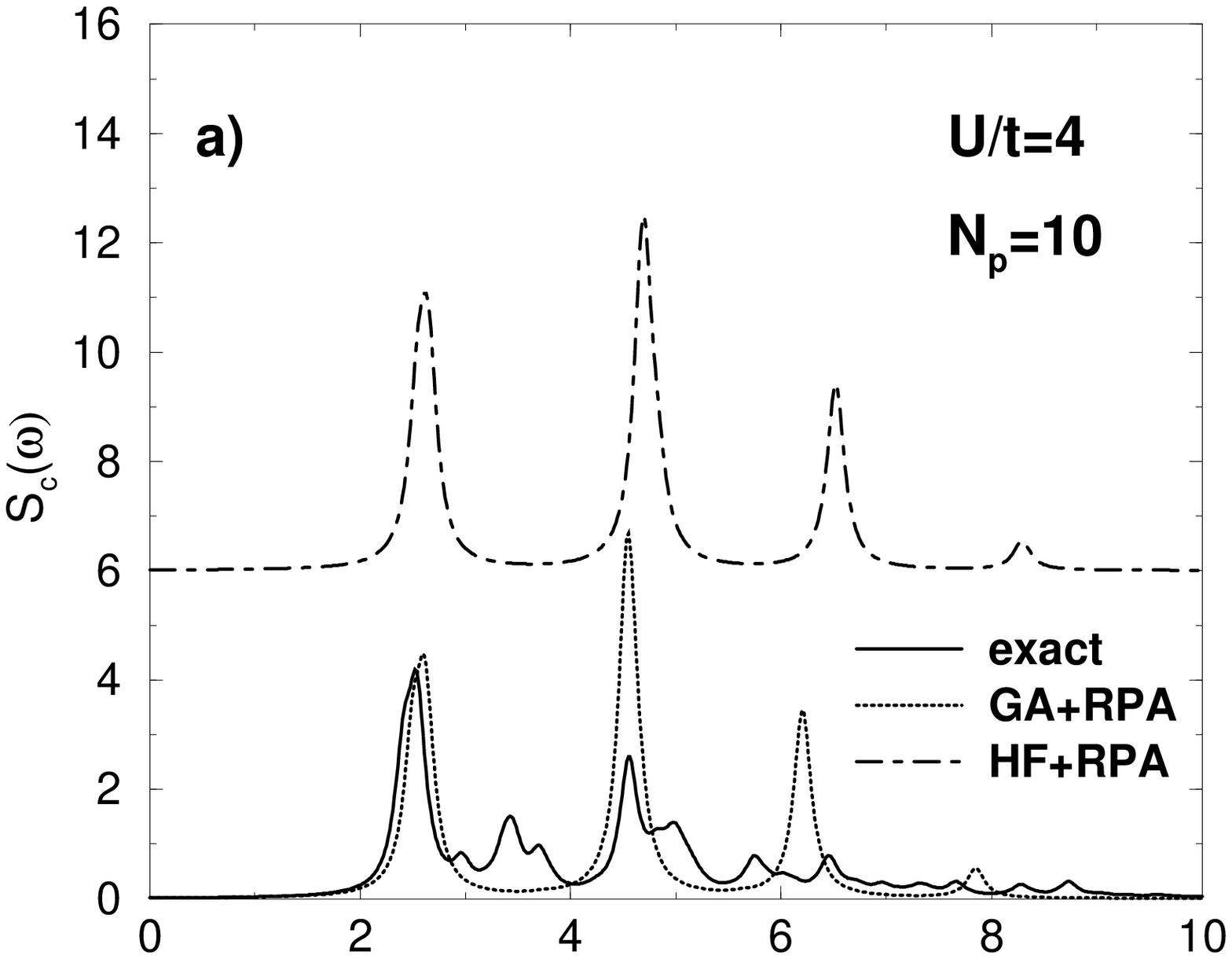}
\includegraphics[width=7.5cm,clip=true]{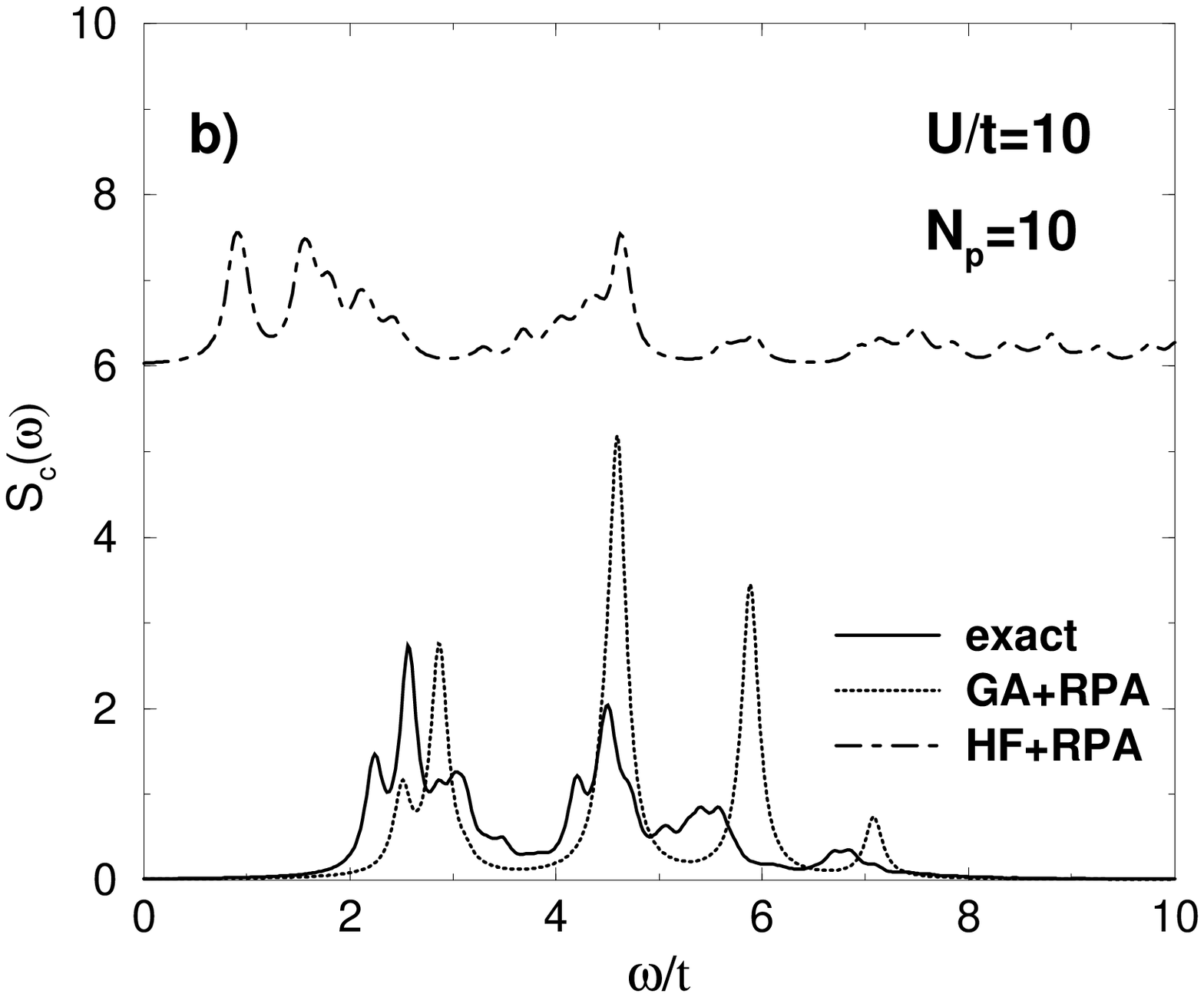}
\caption{Charge correlation function $S_c(\omega)$ for the
$4\times 4$ Hubbard model and 10 particles in case of $U/t=4$ (a)
and $U/t=10$ (b).}
\label{fig:44}
\end{figure}

We finally would like to demonstrate the importance of 
an accurate mean-field solution for the quality of the RPA excitation spectrum.
For this purpose Fig.~\ref{fig:44}, shows S$_c(\omega)$ for a
$4\times 4$ system and 10 particles. In the case of $U/t=4$
[Fig.~\ref{fig:44} (a)] HF+RPA and GA+RPA give a good approximation
to the lowest
excitation, both with respect to the oscillator strength and the energy.
For $U/t=4$ and 10 particles (corresponding to a closed shell configuration)
the underlying mean-field solution are homogeneous with respect to the
charge and do not show SDW order. However, whereas the homogeneous GA solution
remains a stable saddle-point also for large values of $U/t$, the HF solution
becomes instable with respect to a disordered charge and spin texture.
This obviously has dramatic consequences for the dynamical
properties as shown in  Fig.~\ref{fig:44}(b).
In fact, the GA+RPA spectrum still shows a remarkable agreement
with the exact solution in contrast to the RPA on top of
the inhomogeneous HF solution.

Notice that the GA+RPA oscillator
strength for the charge-charge correlations is distributed in a
much better way
than for the optical conductivity (see Fig.~\ref{fig:oc1}).
To some extent, this can be attributed to the constraints imposed
by the sum rules. Both correlation functions satisfy a sum rule
which involves the kinetic energy on the right hand side
[Eqs.~(\ref{fsum}) and~(\ref{SR2})], however, in case of the optical
conductivity the high-energy states contribute
much less to the sum rule than in the case of $S_{c}(\omega)$ due
to the $\omega$ factor. Therefore, the high-energy part of
$S_{c}(\omega)$ is much more constrained to be accurate than the
high-energy part of $\sigma$. This argument assumes that at least the
excitation energies are approximately correct, otherwise one
can satisfy the sum rule by a compensation between the incorrect
excitation energies and the incorrect matrix elements as
it occurs within the bare GA for the charge correlation function
[Fig.~\ref{fig:n14}(b)].

Finally, it is remarkable that the GA+RPA dynamical correlation functions
perform rather well even at energies much larger than the charge
gap (Figs.~\ref{fig:n14},~\ref{fig:44u10} and~\ref{fig:44})  where the antiadiabatic
condition Eq.~(\ref{AAC}) is not expected to hold. To some extent
this may be due to the constraints imposed by sum rules.
It shows that at least for some correlation functions (charge-charge)
the domain of applicability of our theory may be much wider than expected.

\section{Conclusion} \label{conc}
In summary, we have presented a method for the calculation
of dynamical correlation functions in the Hubbard model,
based on the Gutzwiller approximation.
Our method obeys well-behaved sum rules and
we have demonstrated that it is suitable for practical computations:
excitation energies compare
remarkably better with exact diagonalization results
than the related HF+RPA approach.
Moreover, since the performance of any RPA computation
crucially depends on the quality of the underlying mean-field solution
we conclude from our analysis that the GA provides a much better
starting point for this purpose than HF.

The dynamical matrix has been calculated as a quadratic expansion
of the GA energy functional in the densities, which allows us to
construct the RPA eigenvectors and eigenvalues. Essential
assumption for carrying out this expansion is the antiadiabaticity
condition Eq.~(\ref{AAC}). Despite the fact that charge and double
occupancy dynamics seem to be governed by different time scales it
would be desirable to relax the antiadiabatic approximation and to
treat the double occupancy dynamics explicitly.
This is in principle possible via the
Kotliar-Ruckenstein
slave-boson scheme, however, attempts in this direction render difficult due
to the hopping factor expansion.

Compared to numerical methods~\cite{LANCZOS,MONTE} our approach
can be pushed to much larger systems. Our experience on modeling
real data~\cite{LOR,GOETZ2} is that often  finite size effects are
more severe than the inaccuracies introduced by mean-field+RPA
approaches. A more recent approach for dynamical properties
consists of mapping the problem onto quantum impurity models
(dynamical mean-field theory) which becomes exact in the limit of
large dimensions.~\cite{INFDIM} This has enormously increased our
understanding of these systems. However, on making the limit of
large dimensions important parts of the physics are lost. For
example all acoustic like collective behavior, like spin waves
disappears. On the other hand, these collective effects are
naturally captured in our approach.

The GA+RPA formalism can be also applied to multiband Hubbard
models, which are relevant for a more quantitative analysis of
excitations in the cuprate high-T$_c$ superconductors. It has been
shown recently that the GA provides an excellent starting point to
describe the physics of stripes in cuprates including the behavior
of inconmensurability, chemical potential and some transport
experiments with doping.~\cite{LOR3}  In this context it is very
important to make an RPA analysis on top of GA states since within
the HF approximation one obtains a ground state which does not
correspond to experiment. Work in this direction is in progress.

\acknowledgments
G.S. acknowledges financial support from the
Deutsche Forschungsgemeinschaft. The authors  acknowledge  hospitality and
support from the Dipartimento
di Fisica of Universit\`a di Roma ``La Sapienza'' where part of this work
was carried out and J.L. acknowledges hospitality at the Abdus Salam ICTP
(Trieste).

\appendix
\section{Relationship to Kotliar-Ruckenstein slave boson approach }
\label{kotliar} In the Kotliar-Ruckenstein slave-boson approach to
the Hubbard model~\cite{KR} the original Hilbert space is enlarged
by introducing four subsidiary boson fields $e_{i}$,
$s_{i,\uparrow}$, $s_{i,\downarrow}$, and $d_{i}$ for each site
${\bf R}_i$. These operators stand for the annihilation of empty,
singly occupied states with spin up or down, and doubly occupied
sites, respectively. Since there are only four possible states per
site, these boson projection operators must satisfy the
completeness constraints:
\begin{equation}\label{C1}
e_{i}^{\dagger}e_{i}+\sum_{\sigma}s_{i,\sigma}^{\dagger}s_{i,\sigma}
+d_{i}^{\dagger}d_{i}=1,
\end{equation}
and
\begin{equation}\label{C2}
n_{i,\sigma}=s_{i,\sigma}^{\dagger}s_{i,\sigma}+d_{i}^{\dagger}d_{i}.
\end{equation}
In the saddle-point approximation, all bosonic operators are treated
as numbers and
the resulting effective one-particle Hamiltonian $H^{KR}$
describes the dynamics
of particles where the hopping amplitude between states
($i$,$\sigma$) and ($j$,$\sigma$) is renormalized by a factor
$z^{SB}_{i,\sigma}
z^{SB}_{j,\sigma}$ with
\begin{equation}
z^{SB}_{i,\sigma}=
\left( e_{i}^2+s_{i,-\sigma}^2 \right)^{-1/2} (e_{i}s_{i,\sigma}
+s_{i,-\sigma}d_{i})\left( d_{i}^2+s_{i,\sigma}^2 \right)^{-1/2}.
\end{equation}
The total energy of $H^{KR}$ is given by
\begin{eqnarray}
E^{SB}&=&\sum_{ij,\sigma}t_{ij}z^{SB}_{i,\sigma}z^{SB}_{j,\sigma}
\rho_{ij} + U\sum_{i}d_{i}^{2} \label{EKR},
\end{eqnarray}
which has to be minimized (i) with respect to the bosonic fields
within the constraints Eqs.~(\ref{C1}) and~(\ref{C2}) and (ii)
with respect to ${\rho}$ within the subspace of Slater determinants.

The slave-boson energy functional $E^{SB}$ is a function of $4N$ boson
variables where N is the number of lattice sites. Since these bosons
obey the constraints Eqs.~(\ref{C1}) and~(\ref{C2})
one can eliminate 3N of them which
leads to the Gutzwiller energy $E^{GA}$ when one keeps
the double occupancy variable $D=d^2$ for each lattice site.
Thus the expansions of $E^{SB}$ and $E^{GA}$ are connected
via the transformation
\begin{eqnarray}
\frac{\partial z_{i,\sigma}^{GA}}{\partial D}
 &=& \frac{\partial z_{i,\sigma}^{SB}}{\partial d^2}
- \frac{\partial z_{i,\sigma}^{SB}}{\partial s_{\sigma}^2}
- \frac{\partial z_{i,\sigma}^{SB}}{\partial s_{-\sigma}^2}
+ \frac{\partial z_{i,\sigma}^{SB}}{\partial e^2}, \nonumber \\
\frac{\partial z_{i,\sigma}^{GA}}{\partial \rho_{ii,\sigma}}
&=& \frac{\partial z_{i,\sigma}^{SB}}{\partial s_{\sigma}^2}
- \frac{\partial z_{i,\sigma}^{SB}}{\partial e^2}, \nonumber \\
\frac{\partial z_{i,\sigma}^{GA}}{\partial \rho_{ii,-\sigma}}
&=& \frac{\partial z_{i,\sigma}^{SB}}{\partial s_{-\sigma}^2}
- \frac{\partial z_{i,\sigma}^{SB}}{\partial e^2}, \label{TRAFO}
\end{eqnarray}
and the derivatives have to be taken at the saddle-point. Upon
inserting this transformation in Eqs.~(\ref{ZZ1}) and~(\ref{ZZ2})
leads to an analogous energy expansion than Eq.~(\ref{eq:e2rho})
but now within the KR scheme. For paramagnetic solutions this
corresponds to the analysis done in Ref.~\onlinecite{LAVAGNA}.

\end{document}